\documentclass[twocolumn,aps,prl,showpacs,preprintnumbers,amsmath,amssymb]{revtex4}

\usepackage{epsfig}
\usepackage{graphicx}
\usepackage{dcolumn}
\usepackage{bm}

\renewcommand{\dag}{^{\dagger}}
 
\newcommand{\la}{\langle}
\newcommand{\ra}{\rangle}
 
\newcommand{\al}{_{\alpha}}
\renewcommand{\ap}{_{\alpha'}}
\newcommand{\alal}{_{\alpha,\alpha'}}
\newcommand{\apal}{_{\alpha',\alpha}}
 
\newcommand{\sx}{\sigma_x}
\newcommand{\sy}{\sigma_y}
\newcommand{\sz}{\sigma_z}

\newcommand{\De}{\Delta}

\def\gapp{\lower.35em\hbox{$\stackrel{\textstyle>}{\sim}$}}
\def\lapp{\lower.35em\hbox{$\stackrel{\textstyle<}{\sim}$}}

\begin{document}
\title{
Equilibrium Correlation Functions of the Spin-Boson Model with Sub-Ohmic Bath
}
\author{T.~Stauber and A.~Mielke}
\affiliation{Institut f\"ur Theoretische Physik, Ruprecht-Karls-Universit\"at Heidelberg, Philosophenweg 19, D-69120 Heidelberg, Germany
}
\date{\today}

%%%%%%%%%%%%%%%%%%%%%%%%%%%%%%%%%%%%%%%%%%%%%%%%%%%%%%%%%%%%%%%%%%%%%%%%%%%%%
\begin{abstract}
The spin-boson model is studied by means of flow equations for Hamiltonians. Our truncation scheme includes all coupling terms which are linear in the bosonic operators. Starting with the canonical generator $\eta_c=[H_0,H]$ with $H_0$ resembling the non-interacting bosonic bath, the flow equations exhibit a universal attractor for the Hamiltonian flow. This allows to calculate equilibrium correlation functions for super-Ohmic, Ohmic and sub-Ohmic baths within a uniform framework including finite bias. 
\end{abstract}
%%%%%%%%%%%%%%%%%%%%%%%%%%%%%%%%%%%%%%%%%%%%%%%%%%%%%%%%%%%%%%%%%%%%%%%%%%%%%
%
%
%
%%%%%%%%%%%%%%%%%%%%%%%%%%%%%%%%%%%%%%%%%%%%%%%%%%%%%%%%%%%%%%%%%%%%%%%%%%%%%
%%%%%%%%%%%%%%%%%%%%%%%%%%%%%%%%%%%%%%%%%%%%%%%%%%%%%%%%%%%%%%%%%%%%%%%%%%%%%
%%%%%%%%%%%%%%%%%%%%%%%%%%%%%%%%%%%%%%%%%%%%%%%%%%%%%%%%%%%%%%%%%%%%%%%%%%%%%
%%%%%%%%%%%%%%%%%%%%%%%%%%%%%%%%%%%%%%%%%%%%%%%%%%%%%%%%%%%%%%%%%%%%%%%%%%%%%
%
\maketitle

The spin-boson model, consisting of a spin-1/2-system coupled to a bath of non-interacting harmonic oscillators, is the simplest non-trivial model that captures the essential influence of dissipation on a quantum mechanical system \cite{Leg87,Wei99}. The physical behavior of the system is determined by the low-energy modes of the bosonic bath. Modeling the interaction such that one recovers the classical phenomenological equation of motion with a frictional force proportional to the velocity, one speaks of Ohmic dissipation. At zero temperature the system then undergoes a continuous transition from delocalization to localization at a critical interaction strength. If the low-energy modes are suppressed or enhanced relative to the Ohmic bath one speaks of super-Ohmic or sub-Ohmic dissipation. For super-Ohmic dissipation, the effect of the bath on the system is generally that of renormalizing the system parameters. For sub-Ohmic baths, the system has the tendency to localize. 

Physical systems mostly exhibit super-Ohmic or Ohmic dissipation. Obvious realizations for sub-Ohmic baths are not known and the only known results are valid for high-temperatures and small time scales \cite{Wei99}. They are e.g obtained using Monte Carlo simulations \cite{Egg92}. The lack of results for long time scales at zero temperature is mainly due to technical difficulties since the sub-Ohmic coupling represents a relevant perturbation in the language of the renormalization group \cite{Bac95}.

%Matters have changed with the upcoming interest in qubits embedded in a solid-state matrix \cite{Mak01}. The advantage of this implementation of a quantum computer is the possibility to easily scale up the number of qubits and couple and manipulate them through external fields using nanotechnology. The apparent deficit compared to quantum optical qubits is the inherent coupling to a dissipative environment. But investigations on the influence of Ohmic dissipation within the spin-boson model have shown that there are no principal restrictions to quantum coherence as long as the qubit is operated at an appropriate working point \cite{Mak01}. Nevertheless, recent experiments indicate another source of dissipation which is given by $1/f$ noise \cite{Nak02}. Whether or not this noise term can be modeled by a bath consisting of harmonic os

With the upcoming interest in quantum computing based on solid state qubits, investigations on different types of dissipative systems have emerged \cite{Mak01,Ced00,Ngu01} and there are e.g suggestions to model $1/f$ noise found in recent experiments \cite{Nak02} by a limiting sub-Ohmic bath \cite{Shn02}. The widespread believe that there is total suppression of quantum tunneling at zero temperature for any sub-Ohmic bath independent of the coupling strength would pose a serious problem to quantum computing within a solid-state matrix due to $1/f$ noise. But results based on the flow equation approach \cite{Keh96} and on exact mathematical techniques \cite{Spo85} indicate that there is a first order phase-transition from delocalization to localization at some finite coupling constant. To our knowledge, spectral properties which are valid on all energy scales have not been presented yet. 
%This is important in order to thoroughly discuss the transition from coherent to non-coherent tunneling, which sets the relevant benchmark for future implementations of quantum computers.  

In this letter, we will investigate the spin-boson model by means of flow equations \cite{Weg94} since this approach is not confined to any particular bath type. We are thus able to treat all environments on the same footing. One crucial difference to prior works on the method applied to the spin-boson model \cite{Keh97} is, that we allow all coupling terms which are linear in the bosonic bath operators to be generated. Furthermore, we set up flow equations that exhibit universal asymptotic behavior which assures a systematic decoupling procedure where the low-energy modes are decoupled last. In view of the manipulation of qubits, we also treat the spin-boson model with finite bias. This gives rise to distinguished bosonic modes and it will also be necessary to vary the unitary representation of the initial Hamiltonian.

%
%%%%%%%%%%%%%%%%%%%%%%%%%%%%%%%%%%%%%%%%%%%%%%%%%%%%%%%%%%%%%%%%%%%%%%%%%%%%%
%%%%%%%%%%%%%%%%%%%%%%%%%%%%%%%%%%%%%%%%%%%%%%%%%%%%%%%%%%%%%%%%%%%%%%%%%%%%%
%%%%%%%%%%%%%%%%%%%%%%%%%%%%%%%%%%%%%%%%%%%%%%%%%%%%%%%%%%%%%%%%%%%%%%%%%%%%%
%%%%%%%%%%%%%%%%%%%%%%%%%%%%%%%%%%%%%%%%%%%%%%%%%%%%%%%%%%%%%%%%%%%%%%%%%%%%%
%
We start our discussion with the reflection-symmetric model where the initial Hamiltonian is given by
\begin{eqnarray}
\nonumber
    H=\frac{\De}{2}\sigma_z  
        +\omega_{\alpha}b_{\alpha}\dag b\al 
	+\sigma_x\frac{\lambda_{\alpha}^x}{2}(b\al+b\al\dag)\quad.
\end{eqnarray}
Notice that summation over the bath modes $\alpha$ is implied throughout this work. The objective is to evaluate the equilibrium correlation function $C(t)=\la\{\sx(t),\sx\}\ra /2$.

The scale of the isolated system shall be given by $\De_0=\De(\ell=0)$.
The coupling constants $\lambda\al^x$ are as usual parametrized in terms of the spectral coupling function $J(\omega)\equiv(\lambda\al^x)^2\delta(\omega-\omega\al)=2\alpha K^{1-s}\omega^s f(\omega/\omega_c)$ with the coupling strength $\alpha$, the bath cutoff $\omega_c$ and the cutoff function $f(x)$. The different bath types are characterized by $s$, where $s=1$ stands for Ohmic baths and $s>1$ and $s<1$ for super- and sub-Ohmic baths, respectively. Non-Ohmic baths also give rise to an additional energy scale $K$.  

The Hamiltonian is now subjected to a series of infinitesimal transformations which can be written in differential form as $\partial_\ell H=[\eta,H]$, $\ell$ denoting the flow parameter and resembling the square of the inverse energy scale which is being decoupled \cite{Weg94}.
The anti-Hermitian generator $\eta$ is first chosen to be canonical, i.e. $\eta_c=[H_0,H]$ \cite{Weg94}, with $H_0=\omega_{\alpha}b_{\alpha}\dag b\al$. This leads to the trivial fixed point Hamiltonian $H(\ell=\infty)=H_0$. But $\eta_c$ generates a new coupling term which is linear in the bosonic operators and given by $H_{new}=i\sy\frac{\lambda\al^y}{2}(b\al-b\al\dag)$. This term is included in the flow of the Hamiltonian and also contributes to the canonical generator. 

Moreover, the commutator $[\eta_c,H]$ contains coupling terms which are bilinear in the bosonic operators. These terms are not included in the Hamiltonian flow explicitly, but we truncate the Hamiltonian after coupling terms linear in the bosonic operators.
Yet, we consider bilinear coupling terms adequately by introducing a new term to the generator. Doing so, we are guided by the exact solution of the dissipative harmonic oscillator via flow equations \cite{Keh97}. The additional term of the generator is given by $\eta_{new}=\eta\alal(b\al+b\al\dag)(b\ap-b\ap\dag)$, where the parameters $\eta\alal$ are to be chosen such that the bilinear terms are not being generated. This cancellation cannot be done exactly, but we will neglect operators which are normal ordered with respect to the ground-state of the one-particle Hamiltonian $H^p=\frac{\De}{2}\sigma_z$ and the bath Hamiltonian $H_0$, respectively. The parameters $\eta\alal$ for $\alpha\neq\alpha'$ are then given by ($\eta_{\alpha,\alpha}=0$)
\begin{eqnarray}
\label{eta_alal}
\eta\alal=\frac{(\omega\al^2+\omega\ap^2)\lambda\al^x\lambda\ap^y
+2\omega\al\omega\ap\lambda\ap^x\lambda\al^y}{2(\omega\al^2-\omega\ap^2)}\quad.
\end{eqnarray}
The renormalization of the bath modes $\omega\al$ vanishes in the thermodynamic limit and will be neglected. For the tunnel-matrix element we obtain the flow equation $\partial_\ell \De=-2\omega\al\lambda\al^x\lambda\al^y$. The flow of the coupling constants is described by the following non-linear differential equations:
\begin{eqnarray}
\nonumber
&\partial_\ell \lambda\al^x=-\omega\al^2\lambda\al^x+\De\omega\al\lambda\al^y+2\eta\alal\lambda\ap^x&\\\label{couplingflow}
&\partial_\ell \lambda\al^y=-\omega\al^2\lambda\al^y+\De\omega\al\lambda\al^x-2\eta\apal\lambda\ap^y&
\end{eqnarray}

In order to calculate correlation functions the observable has to be subjected to the same sequence of infinitesimal transformations as the Hamiltonian.
The truncated flow of the observable shall be given by $\sx(\ell)=h^x(\ell)\sx+\sz\chi\al^z(\ell)(b\al+b\al\dag)$. The flow equations $\partial_\ell\sx=[\eta,\sx]$ are closed according to the same normal ordering scheme as above and thus read
\begin{eqnarray}\nonumber
\partial_\ell h^x=\omega\al\lambda\al^y\chi\al^z\quad,\quad\partial_\ell \chi\al^z=-h^x\omega\al\lambda\al^y+2\eta\alal\chi\ap^z\quad.
\end{eqnarray}

\begin{figure}[t]
  \begin{center}
    \epsfig{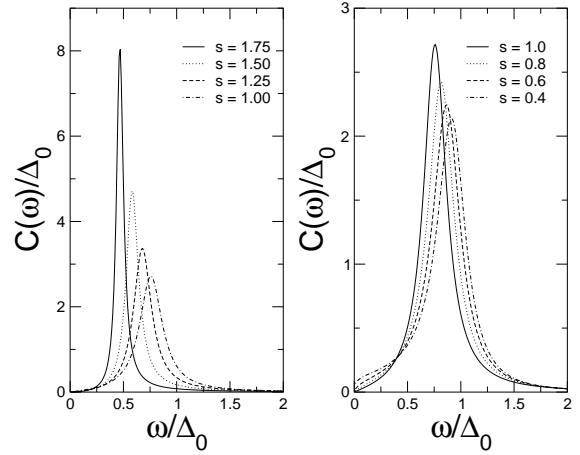}
    \caption{The spectral function $C(\omega)$ for $J(\omega)=2\alpha K^{1-s}\omega^s\Theta(\omega_c-\omega)$ with $\alpha=0.1$, $\omega_c/\De_0=10$, $K/\De_0=1$, and $\epsilon=0$ for various bath type $s$.}
\label{Abb1}
\end{center}
\end{figure}

Characteristic for dissipative systems is that the one-particle parameter $h^x\to0$ for $\ell\to\infty$, i.e. the spectral weight of the system dissipates into the bath \cite{Keh97}. The correlation function $C(t)=\int_0^\infty d\omega e^{i\omega t}C(\omega)$ is then determined by the spectral function $C(\omega)=(\chi\al^z(\ell=\infty))^2\delta(\omega-\omega\al)$ which is obtained by numerical integration of the flow equations. 
In Fig. \ref{Abb1}, $C(\omega)$ is shown for various bath type $s$. The parameters of the spectral coupling function $J(\omega)$ are chosen as $\alpha=0.1$, $\omega_c/\De_0=10$, $K/\De_0=1$. The spectral functions yield the correct low-energy behavior $C(\omega)\propto\omega^s$ for $\omega/\De_0\ll1$ which clearly shows the non-perturbative nature of our approach. We will come back to this point later.  

For the numerical implementation of the flow equations, we employed a linear energy spacing, which allows for an easy evaluation of the principal value integral arising from $\eta\alal$ of Eq. (\ref{eta_alal}), and the sharp cutoff function $f(x)=\Theta(1-x)$. The results for low energies do not depend on the specific choice of $f(x)$ nor on the energy discretization. For larger $\omega_c$ a logarithmic energy spacing is needed which yields equivalent results.
 
\begin{figure}[t]
  \begin{center}
    \epsfig{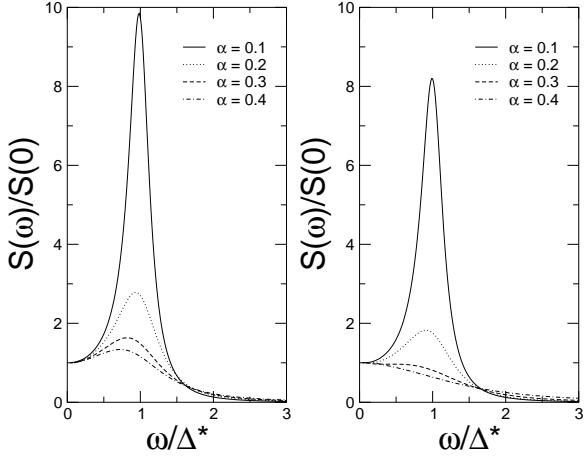}
    \caption{The spectral function $S(\omega)$ for $J(\omega)=2\alpha K^{1-s}\omega^s\Theta(\omega_c-\omega)$ with $\omega_c/\De_0=10$, $K/\De_0=1$, $\epsilon=0$, Ohmic bath $s=1$ (left hand side) and sub-Ohmic bath $s=0.8$ (right hand side) for various coupling strength $\alpha$.}
\label{Abb2}
\end{center}
\end{figure}

In Fig. \ref{Abb2}, the normalized spectral function $S(\omega)\propto C(\omega)/\omega^s$ is shown for various coupling strength $\alpha$ where $\omega=\De^*$ denotes the maximum of $C(\omega)$, i.e. $\De^*\approx\Delta_0(\Delta_0/\omega_c)^{\alpha/(1-\alpha)}$ for Ohmic baths. The parameters of $J(\omega)$ are chosen as $\omega_c/\De_0=10$, $K/\De_0=1$, $s=1$ (left hand side) and $s=0.8$ (right hand side), respectively.
The transition from coherent to incoherent tunneling occurs at the coupling strength $\alpha^*$ where the resonance at $\omega=\De^*$ disappears. We obtain $\alpha^*${\gapp}$0.4$ for an Ohmic bath and $\alpha^*\approx0.3$ for a sub-Ohmic bath with $s=0.8$. 

Our truncation scheme is valid for small and intermediate coupling strengths $\alpha\lapp0.4$, and for Ohmic baths our approach yields results similar to those obtained by Costi and Kieffer by means of the numerical renormalization group (NRG) \cite{Cos96}. But since NRG is based on the mapping of the spin-boson model onto the anisotropic Kondo model it cannot be used for sub-Ohmic baths.

%Our result for Ohmic dissipation is in contrast to $\alpha^*\approx0.33$ obtained by Costi and Kieffer by means of the numerical renormalization group (NRG) after having mapped the spin-boson model onto the anisotropic Kondo model \cite{Cos96}. The discrepancy is due to different bath cutoffs since the mapping is exact only in the limit of infinite bath-cutoff and also only possible for Ohmic baths \cite{Leg87}, which explains the absents of results for sub-Ohmic baths.  

We will discuss our approach in more detail, now. We have chosen the diagonal Hamiltonian to be $H_0=\omega\al b\al\dag b\al$. This choice indicates that $\De(\ell)\to0$ as $\ell\to\infty$, i.e. the fixed point Hamiltonian is scale-independent and thus universal. With no asymptotic scale present, the flow equations will systematically first decouple high-energy modes and finally low-energy modes. This decoupling procedure is passed on to the observable flow, i.e. the spectral function $C(\omega)$ is first determined at high energies and the trivial low-frequency behavior of the system is calculated at last. Moreover, the spectral function around the resonance $\omega\approx\De^*$ is determined by the stable flow of intermediate $\ell\approx(\De^*)^{-2}$, away from the asymptotic regime of decoupling quasi-degenerate states. Another advantage of letting the tunnel-matrix element $\De$ renormalize to zero is that the fixed point Hamiltonian $H_0$ possesses the correct spectrum, consisting of a non-degenerate ground-state and a gapless continuum.

The flow equations can numerically be integrated only up to $\ell^*\approx(\delta\omega)^{-2}$ with $\delta\omega$ denoting the linear energy spacing. Determining the asymptotic behavior will thus specify the spectral function $C(\omega)$ in the low-energy regime which is not accessible by numerical integration. A first approximation for the asymptotic flow is given by $\lambda\al^x\approx\lambda\al^y$, which is consistent with (\ref{couplingflow}). 
Introducing the $\ell$-dependent spectral coupling function $J(\omega,\ell)\equiv(\lambda\al^y(\ell))^2\delta(\omega-\omega\al)$, the asymptotic flow equations are solved by the ansatz $\De(\ell)\to a\ell^{-1/2}$ and $J(\omega,\ell)\to\ell^{-1/2}J(\omega\sqrt{\ell})$. This leads to the following non-local differential equation for $J(y)$ with $y\equiv\omega\sqrt{\ell}$:
\begin{eqnarray}\nonumber
\partial_y J(y)=-4(y-a)J(y)+8J(y)\int_0^\infty dy'\frac{J(y')}{y-y'}+t\frac{J(y)}{y},
\end{eqnarray}
where $t=1-4\int_0^\infty dy'J(y')$ and the boundary condition $a=4\int_0^\infty dyyJ(y)$.

The asymptotic behavior of the observable flow is described by a similar ansatz, i.e. $h^x(\ell)\to b\ell^{-1/4-s'/4}$ and $S(\omega,\ell)\equiv\lambda\al^y(\ell)\chi\al^z(\ell)\delta(\omega-\omega\al)\to\ell^{-1/4-s'/4}S(\omega\sqrt{\ell})$. The non-local differential equation for $S(y)$ is similar to the one for $J(y)$ \cite{StaU}. Defining the $\ell$-dependent spectral function $C(\omega,\ell)\equiv(\chi\al^z(\ell))^2\delta(\omega-\omega\al)$, we obtain for $\ell^*$ being in the asymptotic regime $C(\omega)-C(\omega,\ell^*)=\omega^{s'}I(y^*)$ with $y^*=\omega\sqrt{\ell^*}$ and
\begin{eqnarray}
\nonumber
I(y^*)&\equiv&4\int_{y^*}^{\infty}dy y^{1-s'}\big[-b\frac{S(y)}{y}+2\frac{S(y)}{y}\int_0^\infty dy'\frac{S(y')}{y-y'}\\\label{Asymp}
&-&\frac{S(y)}{y^2}\int_0^\infty dy'S(y')\big]\quad.
\end{eqnarray}
For an initial spectral coupling function $J(\omega)\propto\omega^s$, the numerical analysis yields $s\approx s'$ independent of the coupling strength. The asymptotic scale-invariant flow thus induces the correct long-time behavior of the correlation function $C(\omega)$.

There are two criteria for the assessment of the quality of the results. One is the normalization of the spectral function $C(\omega)$, i.e. $\int_0^\infty d\omega C(\omega)=1$. The second is the Shiba relation $\lim_{\omega\to0}C(\omega)/\omega^s=2\alpha K^{1-s}\int_0^\infty C(\omega)/\omega$ \cite{Sas90}. 
The numerical results show that the spectral sum rule is fulfilled within less than 0.1 \% relative error. The interpretation of the Shiba relation is not as obvious. From (\ref{Asymp}) we formally obtain $\lim_{\omega\to0}C(\omega)/\omega^s=I(0)$, which is not defined since the term $\int_{y^*}^\infty dyS(y)/y^{1+s'}$ diverges for $y^*\to0$. But this is not the correct result because the proper order how to perform the limites is first $\ell\to\infty$ and then $\omega\to0$. 
The fact that the asymptotic and the thermodynamic limit do, in general,  not commute is well-known \cite{Weg94,Keh97}. 

Performing the limits correctly, the left and right hand side of the Shiba relation can be determined. This yields satisfactory results with relative errors of around $10\%$ for $\alpha\lapp 0.2$ and $s\gapp0.8$. For larger $\alpha$ and lower $s$ the relation still holds with a maximal relative error of around $30\%$. Approximation schemes like NIBA or numerical methods based on Monte Carlo cannot be used to verify the Shiba relation since they fail to predict the correct long-time behavior. Results from NRG \cite{Cos96} for the Ohmic bath have a maximal relative error of around $10\%$.

Applying an external bias, the Hamiltonian flow as well as the operator flow is not confined to one or two basic interaction terms anymore. In this case, we allow all coupling terms which are linear in the bosonic operators with real coefficients to be generated. The full truncated Hamiltonian thus reads 
\begin{eqnarray}\nonumber
H=-\frac{\De}{2}\sigma_x+\frac{\epsilon}{2}\sigma_z  
        +\omega_{\alpha}b_{\alpha}\dag b\al 
	+\sigma_j\frac{\lambda_{\alpha}^j}{2}x\al-\sigma_y\frac{\lambda_{\alpha}^y}{2}p\al\quad,
\end{eqnarray}
with $x\al\equiv(b\al+b\al\dag)$ and $p\al\equiv -i(b\al-b\al\dag)$. Notice that summation over $j=e,x,z$ with $\sigma_e\equiv1$ is implied from now on.

\begin{figure}[t]
  \begin{center}
    \epsfig{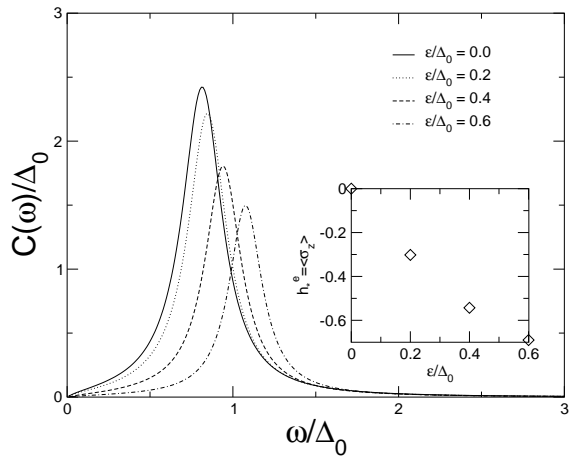}
    \caption{The spectral function $C(\omega)$ for $J(\omega)=2\alpha K^{1-s}\omega^s\Theta(\omega_c-\omega)$ with $\alpha=0.1$, $\omega_c/\De_0=10$, $K/\De_0=1$, and sub-Ohmic bath $s=0.8$ for various bias $\epsilon$. At $\omega=0$ there are $\delta$-peaks with weight $(h_*^e)^2$ (Indent).}
\label{Abb3}
\end{center}
\end{figure}

In the above representation, the considered equilibrium correlation function is defined by $C(t)=\langle\{\sigma_z(t),\sigma_z\}\rangle/2$. The extended truncated observable flow is thus given by
\begin{eqnarray}\nonumber
    \sigma_z(\ell)=h^j(\ell)\sigma_j+\sigma_j\chi_{\alpha}^j(\ell)x\al-\sigma_y\chi_{\alpha}^y(\ell)p\al\quad,
\end{eqnarray}   
where $\chi\al^e(\ell)=0$ for all $\ell$. It is important that the constant term $h^e$ is generated which indicates reflection-symmetry breaking, i.e. $\langle\sigma_z\rangle=h^e(\ell=\infty)\equiv h_*^e$. Hence, it contributes to the spectral function $C(\omega)$ via a $\delta$-function at $\omega=0$ with weight $(h_*^e)^2$.

In principle, each unitary equivalent initial Hamiltonian would yield the same physical result. Matters change as soon as approximations are involved since the neglected terms become differently important depending on the representation of the initial Hamiltonian. Investigating the system with finite bias $\epsilon$ we thus introduce another parameter $\theta$ which indicates an initial shift of the bosonic operators, i.e $b\al\to b\al+\theta\frac{\lambda\al^z}{2\omega\al}$. We choose $\theta$ such that the sum rule $\langle\sigma_z^2(\ell)\rangle=1$ is fulfilled best for all $\ell$. We also diagonalize the initial Hamiltonian with respect to the two-dimensional Hilbert space before we apply the flow equations. 

To set up the flow equations for the extended model, we proceed in the same way as in the case of the symmetric one. Only, the normal ordering procedure now has to be defined with respect to distinguished, $\ell$-dependent modes $\bar{b}\al\equiv b\al+\delta\al/2$ with 
\begin{eqnarray}
\nonumber
\delta\al=\langle\sigma_j\rangle\frac{\lambda\al^j}{\omega\al}\quad,  
\end{eqnarray}
where the one-particle expectation values are evaluated with respect to the effective one-particle Hamiltonian $H^p=-\frac{\De'}{2}\sigma_x+\frac{\epsilon'}{2}\sigma_z$ with $\De'=\De+\lambda\al^e\lambda\al^x/\omega\al-\lambda\al^y\lambda\al^z/\omega\al$ and $\epsilon'=\epsilon-\lambda\al^e\lambda\al^z/\omega\al-\lambda\al^x\lambda\al^y/\omega\al$. The distinguished modes are derived by quadratic completion and suitable factorization, see Ref. \cite{StaD} for details and extensions.

In Fig. \ref{Abb3}, the spectral function $C(\omega)$ for initial coupling $J(\omega)$ with $\alpha=0.1$, $\omega_c/\De_0=10$, $K/\De_0=1$, and sub-Ohmic bath $s=0.8$ is shown for various bias $\epsilon$. The initial shift of the bosonic operators is taken to be $\theta=\epsilon/\De_0$. This yields a spontaneous breaking of the reflection symmetry which is favorable for the system flow as well as for the observable flow. In the indent, the expectation value $\langle\sigma_z\rangle=h_*^e$ is shown. Similar results are obtained for super-Ohmic and Ohmic baths.

In summary, we have established a flow equation scheme which yields reliable results for a large regime of the parameter space of the spin-boson model with essentially new results for sub-Ohmic baths and finite bias. Moreover, we are not confined to specific cutoff functions which might be crucial in view of environments based on a structured bath \cite{Wil02}.

We wish to thank F. Wegner and S.K. Kehrein for valuable discussions. This work was partially supported by the Deutsche Forschungsgemeinschaft.

\end{document}